# Substrate mediated nitridation of niobium into superconducting Nb$_2$N thin films for phase slip study

*Bikash Gajar[1,2], Sachin Yadav[1,2], Deepika Sawle[1,2], Kamlesh K. Maurya[1,3], Anurag Gupta[1,2], R.P. Aloysius[1,2], and Sangeeta Sahoo*[1,2]*

[1] *Academy of Scientific and Innovative Research (AcSIR), AcSIR Headquarters CSIR-HRDC Campus, Ghaziabad, Uttar Pradesh 201002, India.*

[2]*Electrical & Electronics Metrology Division, National Physical Laboratory, Council of Scientific and Industrial Research, Dr. K. S Krishnan Road, New Delhi-110012, India.*

[3]*Indian Reference Materials Division, National Physical Laboratory, Council of Scientific and Industrial Research, Dr. K. S Krishnan Road, New Delhi-110012, India.*

*\*The correspondence should be addressed to S.S. (Email: sahoos@nplindia.org)*

## Abstract:

Here we report a novel nitridation technique for transforming niobium into hexagonal $Nb_2N$ which appears to be superconducting below 1K. The nitridation is achieved by high temperature annealing of Nb films grown on $Si_3N_4$/Si (100) substrate under high vacuum. The structural characterization directs the formation of a majority $Nb_2N$ phase while the morphology shows granular nature of the films. The temperature dependent resistance measurements reveal a wide metal-to-superconductor transition featuring two distinct transition regions. The region close to the normal state varies strongly with the film thickness, whereas, the second region in the vicinity of the superconducting state remains almost unaltered but exhibiting resistive tailing. The current-voltage characteristics also display wide transition embedded with intermediate resistive states originated by phase slip lines. The transition width in current and the number of resistive steps depend on film thickness and they both increase with decrease in thickness. The broadening in transition width is explained by progressive establishment of superconductivity through proximity coupled superconducting nano-grains while finite size effects and quantum fluctuation may lead to the resistive tailing. Finally, by comparing with Nb control samples, we emphasize that $Nb_2N$ offers unconventional superconductivity with promises in the field of phase slip based device applications.

# Introduction:

Superconducting fluctuations (SFs) lead to several interesting quantum phenomena such as superconductor-insulator quantum transition (SIT)[1,2], quantum criticality[3], phase slip (PS) effects[4-7], etc. These quantum phenomena are mainly controlled by sample properties like geometry and dimensions, crystallinity, disorder and inhomogeneity and also by external parameters like electro-magnetic field, driving current, temperature etc. In superconductor (SC)-metal (NM) phase transition probed by temperature dependent resistance [$R(T)$] measurements, a finite transition width is inevitable as it is quite natural for a practical sample to be introduced with disorder and inhomogeneities during the growth and fabrication process. For low dimensional superconductor, inhomogeneity and superconducting fluctuations are the two mostly addressed origins behind wide transition width observed in $R(T)$. Inhomogeneities can be of structural imperfections like chemically impure samples, granularity and grain boundaries etc. and also it can be of geometrical nature such as constrictions and non-uniform edges. The geometrical inhomogeneities should be considered in great detail when a material possesses size dependent $T_c$ variation which may cause broadening in $R(T)$[6]. However, in reduced dimension, the phase fluctuation of the order parameter contributes significantly to the $R(T)$ broadening. Particularly in 1D superconducting nanowire, continuous phase fluctuation leads to finite resistance below $T_c$ by phase slippage of $2\pi$ at the phase slip centers (PSCs). This is known as phase slip (PS) phenomenon which can be established in wide 2D superconducting strips by the appearance of phase slip lines (PSLs).

In the context of SFs, more specifically superconducting phase fluctuations, NbN from the family of niobium-nitrogen based transition metal nitrides (TMNs) has shown its promises for the application in superconducting nanowire based single photon detectors (SNSPD)[8] and quantum phase slip (QPS) based devices[9-11]. However, no other member from this TMN family has been explored at the equal footing in the field of superconductivity. Recently, hard hexagonal ε-NbN has been shown to possess superconducting properties with $T_c$ ~11 K[12]. Among the several stable phases in the niobium-nitrogen based compounds, $Nb_2N$ has recently attracted a growing interests towards its epitaxial growth on SiC substrate[13]. $Nb_2N$ is known mainly for its hardness[14] and a little is known about its superconducting properties[15]. In this study, we report of superconducting properties for $Nb_2N$ thin films that are grown using a novel substrate mediated synthesis process. Here, we present the study of PS phenomenon in quasi 2D disordered granular $Nb_2N$ thin films by means of current-voltage measurements. Compared to the expensive MBE based growth technique[15], we use a very simple route[16] to achieve $Nb_2N$ by high temperature annealing of sputtered grown Nb thin films on $Si_3N_4$/Si (100) substrate. High resolution X-ray diffraction (HRXRD) confirms of $Nb_2N$ as the majority phase along with a minority phase $Nb_4N_5$. Further, $R(T)$ measurements on $Nb_2N$ thin films varying in thickness from 8-16 nm exhibit NM-SC transition at ~ 1K with broad transition width consisting of two distinct regions. While comparing with equivalent Nb control samples grown on $SiO_2$/Si substrate in the same run, the transport measurements for $Nb_2N$ films appear drastically different and unconventional in nature. The zero-field current-voltage characteristics (*IVC*s) are also wide and equipped with intermediate resistive steps which are originated from phase slip lines (PSLs). The transition widths in $R(T)$ and in *IVC*s strongly depend on the film thickness and a wider transition is observed for thinner samples. The resistive transition can be understood from the granular nature of the samples that can be constructed as a collection of randomly distributed nanoscale granules separated by grain boundaries and the two-step $R(T)$ transition characteristics can be explained in terms of local and global superconductivity[17,18]. In this case, the macroscopic superconductivity is established progressively through the Josephson proximity coupling effect among the locally disconnected superconducting grains possessing dimensional dependent distribution of critical temperature. We emphasize that disorder induced inhomogeneity[19], finite size effect (FSE) for quasi zero-dimensional (0D) grains and the phase fluctuation are the main reasons behind the wide transition and resistive tailing. Besides, the appearance of PSLs in the *IVC*s and their evolution with film thickness indicate that disordered superconducting $Nb_2N$ can be of future interests in the field of PS based studies and applications.

## Results:

We present the substrate mediated nitridation technique for Nb to form $Nb_2N$ schematically in Fig. 1(a)-(d). We use $Si_3N_4/Si(100)$ as the substrate where the only source of nitrogen for nitridation is the dielectric $Si_3N_4$. When a $Si_3N_4/Si(100)$ substrate is heated at temperature ~ 820°C under high vacuum, $Si_3N_4$ decomposes into elemental Si and N atoms [20] and the lighter N atoms acquire high kinetic energy and become easily mobile to interact with the deposited metallic film[16]. The decomposition process and the movement of the split Si and N atoms are presented in (b) and (c). The yellow arrows in (c) show the movement of N atoms and the white arrows indicate the position of heavier Si atoms close to Si substrate. The growth of Nb films on $Si_3N_4/Si$ substrate at $(820 \pm 10)$°C by using magnetron sputtering is schematically displayed in Fig. 1(c). After completing the deposition we continue annealing at the same temperature for two hours at pressure ~ 0.5-1.5 x $10^{-7}$ Torr. At this stage, Nb atoms interact with the released N atoms from $Si_3N_4$ and undergo the nitridation process. The chemical reaction during the annealing process is shown in Fig. 1(e) by using the respective crystal structures of $\beta$-$Si_3N_4$, bcc Nb and hexagonal $Nb_2N$.

For comparison, we have deposited Nb films on $Si_3N_4/Si$ (nitride) and $SiO_2/Si$ (oxide) substrates simultaneously and the latter act as control samples. In Fig. 2, we present XRD spectra for 4 representative samples (G1, G2, G3 and G4) with varying thickness and a control sample C1. The thickness values of the samples G1, G2, G3 and G4 are about 80 nm, 40 nm, 17 nm and 12 nm, respectively. The control sample C1 is grown with sample G3 in the same run and they were closely placed during the growth process and are having almost same thickness ~ 17 nm with ± 2 nm variation. The XRD pattern for the control sample reveals majorly the cubic Nb phase. However, some oxide phases for Nb appear too due to high temperature annealing on oxide substrate[21]. For the samples on nitride substrates, majority of the peaks relate to hexagonal $Nb_2N$ phase and consequently, a clear difference from the control sample is evident. All strong peaks related to hexagonal $Nb_2N$ phase are present and a couple of other relatively weak $Nb_4N_5$ peaks appear too. The appearance of elemental Nb peaks depends on the thickness of the films. For example, the two relatively thick samples G1 and G2 show the presence of Nb in addition to its nitride phases. With decreasing thickness the relative amplitudes of Nb peaks get reduced and the nitride phases start to become prominent and finally for G3, almost there is no trace of Nb and the same is true for the thinnest sample G4. Here the peak, at $2\Theta = 38.5°$, corresponds to Nb (110) plane as evident for the control sample and the same is present in the XRD spectra of G1 and G2. However, if we compare the amplitude of this peak in these three samples shown from the bottom to upwards in Fig. 2, we find that the relative amplitude gets reduced in G1 as compared to that in C1 and a further reduction occurs from G1 to G2 for which a very weak peak appears at that position. Now if we move upward to sample G3, we observe a much stronger peak appears at the same position as that of Nb (110) plane along with other strong $Nb_2N$ peaks. Here it is important to note that other Nb peaks disappear completely from the XRD pattern for G3. Incidentally, at *$2\Theta = 38.5°$*, $Nb_2N$ possesses its strongest peak related to (101) plane. As most of the peaks correspond to $Nb_2N$ phase and no other peaks of Nb are present, the afore-mentioned peak certainly indicates $Nb_2N$ (101) plane. Therefore, the majority of the peaks guide towards the formation of $Nb_2N$ phase. Recently, it has also been shown that single phase $Nb_2N$ growth can be achieved in the temperature range 750-850°C which is consistent with our observation[13].

As the substrate induced nitridation for Nb is evident, we call the samples on $Si_3N_4$/Si and $SiO_2$/Si substrates as nitride samples and Nb control samples, respectively. For low temperature transport measurements, the adapted device geometry is shown in the inset of Fig. 3(a). A set of *R(T)* is displayed in Fig. 3 where three nitride samples with varying thickness along with three Nb control samples are presented. The nitride samples B1 [(16 ± 1.6) nm], B2 [(11 ± 2) nm] and B3 [(8 ± 1.4) nm], along with their respective control samples $B1^{Ox}$, $B2^{Ox}$ and $B3^{Ox}$ are grown in three different batches. With thickness, the control samples show noticeable changes in their critical temperature ($T_c$) and normal state resistance ($R_N$) as evident in Fig. 3(a). Here, by lowering thickness, $T_c$ decreases and $R_N$ increases. Now, for the nitride samples, we observe a drastic change in the values for the $T_c$ compared to that of the corresponding Nb samples. Broad metal-superconductor (NM-SC) transition featuring intermediate hump like structure is observed for all the nitride samples. Further, $R_N$ decreases anomalously with decreasing thickness. For better comparison, the normalized resistance $R/R_N$ is presented in Fig. 3(b). The clear differences in the transition width for nitride and oxide samples are evident. For the nitride samples, the NM-SC transition occurs in two steps split by the intermediated hump. Accordingly, we have marked the two transition regions, separated by the horizontal dotted line, and defined the critical temperatures for the first onset temperature ($T_c^{Onset-I}$), the second onset temperature ($T_c^{Onset-II}$), and the superconducting critical temperature ($T_{c0}$) as indicated by the arrows in Fig. 3(b) for the representative sample B3. The transition regions I & II are defined as the regions between $T_c^{Onset-I}$ and $T_c^{Onset-II}$ & $T_c^{Onset-II}$ and $T_{c0}$, respectively. The regions are clearly distinguishable in Fig. 3(c) which shows $R/R_N$ in the reduced temperature scale ($T/T_{c0}$) for nitride samples. Region-I appears differently for different samples whereas, region-II reveals very little change in the overall transition region among the samples. Besides, the relatively higher slope of the transition in Region-II indicates that the superconducting transition in region-II is mainly due to the majority of the phases present in all the nitride samples i.e. the $Nb_2N$ phase. The wide transition in region-I mainly indicates the influence of inhomogeneity including the chemical impurities like the presence of other nitride phases as evident in the XRD spectra[12]. Further, region-I features the widest transition for the thinnest sample B3 and the width gets reduced with increasing the thickness. The overall dependence of the transition regions on film thickness, by means of characteristic critical temperatures, is displayed in Fig. 3(d). It is apparent that the transition width varies strongly with thickness for region-I whereas it remains almost unaffected in region-II. Further, the final step of transition from $T_c^{Onset-II}$ to the superconducting state occurs at lower temperature for thinner sample. Therefore, the thinnest sample (B3) offers the lowest $R_N$ value, the highest $T_c^{Onset-I}$, the lowest $T_c^{Onset-II}$ among the samples under consideration. The characteristic parameters are summarized in a tabular form in the Supporting Material (SM).

Further, we have measured the current-voltage characteristics (*IVC*s) at different temperatures and the *IVC* isotherms are shown in Fig. 4. The *IVC*s for the nitride samples B1, B2 and B3 are shown in Fig.

4(a), (b) and (c), respectively. First, the *IVC*s are wide and hysteretic in nature for all the three samples. Here for clarity we only present the increasing current direction, i.e. up sweep, and the *IVC*s in both up and down sweep direction is shown in the SM. The SC-to-NM transition for B1 is not occurring in a single step for the temperature range from 60 mK to 500 mK and an intermediate resistive step appears. The current-extent for the step decreases with increasing temperature and at about 600 mK, the transition takes place smoothly from SC to NM state without showing any intermediate step. Similarly, intermediate resistive steps emerge for the relatively thinner sample B2 as presented in Fig. 4(b). Interestingly, we observe more than one intermediate steps with increasing slopes that follow each other and converge at the excess current $I_s$. With further reduction of the thickness as in B3, much wider SC-NM transition region consisting of multiple resistive steps in the *IVC*s is observed. For clarity, we have plotted them separately along with down sweeps in Fig. 4(d), where the *IVC*s are shifted in the voltage axis and the starting points correspond to zero-voltage. The steps in the *IVC*s are extended by the dotted lines those meet at the excess current $I_s$. The resistive step like features in between SC and NM states may originate due to phase slip lines (PSLs) in 2D superconducting films[22-24]. However, the extent between the resistive states gets widened by thinning as observed from B1-to-B2-to-B3. Further, the smoothly varying region with finite voltage drop, between $I_{c0}$ and the first resistive transition, indicates the presence of slow moving vortex-antivortex pairs (VAPs) originated at the edges of 2D superconductors [25]. For thinner samples, the aforementioned region in the *IVC*s gets elongated before merging into the phase slip lines at the instability point [23,26]. However, the existence of $I_s$ and increasing slopes for the higher order resistive steps are the signatures of PSLs [9,23,24]. Furthermore, with decreasing thickness, the number of PSLs increases and the whole span between SC-to-NM gets widened. This might be due to the samples becoming more disordered by thinning[13]. Finally, as the current leads are of bigger width than that of the superconducting strips, the lead induced inhomogeneity to cause the resistive steps in *IVC*s can be ignored and hence the PSLs appearing in the *IVC*s can be considered as the analogue of the phase slip centres (PSCs) as appear in 1D nanowire[27,28].

As the switching from SC-to-NM does not occur in a single step, we have multiple characteristic critical currents among which two extreme cases are considered. The first one, defined as the critical current $I_{c0}$, relates to the onset of finite voltage from the SC state and the second one, named as $I_{nm}$, associates with the transition to complete normal state. The retrapping current $I_r$ is defined as the current when the sample starts to transit from its resistive state to the superconducting state in returning direction. The characteristic currents and the sweeping directions are shown by the dotted and the solid arrows, respectively in Fig. 4. The temperature dependence of the characteristic critical currents $I_{c0}$, $I_{nm}$ and $I_r$ for all the three nitride samples along with a couple of Nb control samples on oxide substrate are shown in Fig. 5. First, for all the nitride samples, the retrapping current $I_r$ is higher than the critical current $I_{c0}$, whereas, for the control samples shown in the insets of Fig. 5(a)&(c), the critical current is much higher than the retrapping current which usually occurs when the *IVC*s are hysteretic in nature. However, higher value of retrapping current than the critical current actually indicate towards the weak-links (WLs) formed by the nanostructured grains that might be in resistive state even at temperatures below $T_C$ [24,29]. For the control samples, the span between $I_c$ and $I_r$ increases with lowering the measurement temperature. For particularly thinner nitride samples B2 and B3, $I_{c0}$ and $I_r$ seem to be very close to each other and the variation remains almost unchanged when we reduce the temperature. However, $I_{nm}$ increases strongly and hence the extent in $I_{c0}$ or $I_r$ and $I_{nm}$ gets widened while lowering the temperature. For better comparison, we have plotted in Fig. 5(d) the two extreme critical currents $I_{c0}$ & $I_{nm}$ and retrapping current $I_r$ in one graph. The extent between $I_{c0}$ & $I_{nm}$ is shown by the vertical arrows in the left side. The largest (smallest) width in current is observed for the thinnest (thickest) sample B3 (B1) and this is consistent with the *IVC* isotherms presented in Fig. 4 which shows that the intermediate steps indicating PSLs are much wider and more in numbers for the thinner samples. Therefore, thickness is playing a crucial role to control over the transition region and PS mechanism in the nitride samples.

In order to observe the effect of magnetic field on the *IVC*s, we have measured *IVC*s for sample B1 under perpendicular magnetic field at 60 mK and the same is shown in Fig. 6. As it is obvious, with magnetic field $I_{c0}$ decreases but interestingly, one extra intermediate step appears at 50 mT compared to the zero field *IVC*. Up to 150 mT the intermediate steps are following each other and they appear much more widened than that at zero-field. At 50 mT, the transition to normal state happens at much higher current than the same at zero-field. Another point to note is the curvature of the *IVC*s in the span between $I_{c0}$ and the first resistive step changes from convex to concave under the applied field. In Fig. 4, we observe a similar type of curvature change with reduction in thickness from samples B1 through B2 to B3. The number of resistive steps is also observed to increase in the same sequence from B1 to B2 to B3. Therefore, the effect of magnetic field on the *IVC*s compliments the thinning effect at zero-field[24,30].

## Discussion:

In contrast to Nb control samples, both *R(T)* and *IVC*s show wide transition for nitride samples. The *R(T)* data features two distinct regimes in the transition. The *IVC*s, being hysteretic in nature, showcase intermediate resistive steps which are the signatures of PSLs in wide superconducting films[22,24]. For granular films, often the broadening in *R(T)* is referred to inhomogeneity[19] and finite size effects (FSE) that depend mainly on the grain size and grain boundaries[19]. The topography images (SM), obtained by atomic force microscopy (AFM), reveal of granular nature and the grain size depends on the sample thickness. The average grain size decreases with the thickness and the same is 40 nm, 30 nm, and 23 nm for B1, B2, and B3, respectively. The variation in the grain size is about ± 10 nm for all the samples. Here, the maximum relative variation in the grain size occurs for the thinnest sample B3 which undergoes the widest transition [Fig. 3]. For thinner samples, an enhanced $T_c$ is expected as thinning leads to a shorter mean free path and the broadening in *R(T)* can be due to the enhanced $T_c$ [7,31]. Interestingly for region-I, we have observed that $T_c^{Onset-I}$ increases for reduced film thickness and the thinnest sample B3 offers the highest $T_c^{Onset-I}$. However, region-II does not show strong variation in the transition width among the samples. Here, $T_c^{Onset-II}$ varies in a regular manner i.e. it gets reduced by thinning and the variation is very little particularly in the reduced temperature scale as shown in Fig. 3(c).

In order to have an insight into the NM-SC phase transition as appeared in the measured $R(T)$ characteristics, we present the normalized resistance in the reduced temperature scale for the nitride samples separately in Fig. 7(a). For clarity, different stages of the transition are marked by the vertical dotted lines and the different regions are shaded with different color. The resistive transition can be understood from the granular nature of the samples that can be constructed as a collection of randomly distributed nanoscale granules separated by grain boundaries as shown schematically in Fig. 7(b). Usually for granular films, the two-step NM-SC transition as observed in $R(T)$ data can be explained in terms of local and global superconductivity[17,18,32-34]. Here the differences in the grain size represent the inhomogeneous nature of the samples. With lowering the temperature, individual metallic grains undergo the NM-SC phase transition with a distribution in their transition temperatures. At this stage the superconductivity is established locally at individual granule level which eventually reduces the resistance of the system and hence, a drop in the resistance from the normal state value is expected to appear in the $R(T)$. As observed in Fig. 7(a), resistance starts to drop at $T_c^{Onset-I}$, the temperature at which individual grains started to undergo the phase transition locally. Further lowering the temperature, the drop in resistance continues as more numbers of granules become superconducting and the situation is depicted in Fig. 7(c). As evident in Fig. 7(a) and Fig. 3(c), the temperature extent in region-I varies strongly among the samples that are of different thickness. With further reducing the temperature down to $T_c^{Onset-II}$, the resistive transition takes steeper step and the region-II starts. Here the superconducting nano-grains interconnect to each other through Josephson proximity coupling effect (PE)[35-37]. Through PE, the closely spaced superconducting granules coherently couple together and form superconducting puddles of bigger dimension and the progressive superconductivity is established. In Fig. 7(d), schematically we illustrate the formation of superconducting puddles by combining phase coherent superconducting nano-grains in region-II where the phase coherence among the puddles is impaired by the phase fluctuation. Finally, below $T_{c0}$, the puddles couple coherently altogether to establish the coherent macroscopic SC-state as depicted in Fig. 7(e). In region-II, the relatively sharp resistive transition can be understood by the continuous phase fluctuation among the weakly coupled superconducting puddles and this region can be explained by the Berezinskii-Kosterlitz-Thouless (BKT) transition which relates to the crossover from the fluctuating region-II to the condensed superconducting state. Below the BKT transition temperature $T_{BKT}$, no unbound vortex-antivortex pair exists and long range order is established and ideally it is a zero-ohmic resistive state. The cyan curves in Fig. 7(a) represent the BKT fits to the resistive transition in region-II by using Halperin-Nelson equation[38,39]

$$R(T) = R_0 exp[-b/(T-T_{BKT})^{1/2}] \qquad (1)$$

Where, $R_0$ and $b$ are the material-specific parameters. The selected region of $R(T)$ containing the fitted curves is presented separately in Fig. 7(f) which shows that the measured $R(T)$ can be explained fairly well by the BKT model as given by equation (1).

However as shown in Fig 7(a) &(f), for all the samples close to their $T_{c0}$, the transition from region-II to SC-state accompanies rounded bottom near. For B1, the BKT fit follows very closely to the experimental data and the $T_{BKT}$ appears to be the same as $T_{c0}$ (0.73 K). However, the fits for B2 and B3 deviate near their respective zero-ohmic resistance states and the deviation is more prominent for the thinnest sample B3. Further, the transition to superconducting state features resistive tailing below $T_{c0}$ for all the samples. For clarity, a semi-logarithmic plot of selected region from the measured $R(T)$ is shown in Fig. S6 in SM.

Generally, microscopic inhomogeneities[40], disorder[41], vortex-antivortex movement in 2D[23], quantum fluctuation[4], FSE particularly for the case of nanoscale granular systems[35,40,42,43] are the main reasons behind the rounding of the transition, deviation from the BKT model, resistive tailing and residual resistance at temperature $T \leq T_{c0}$. Inhomogeneity and disorder play very crucial roles in the granular superconductors where disorder can cause a suppression of the PE coupling which establishes the superconductivity[19]. As the thinning makes the samples more disordered it is expected that for the thinnest sample B3, disorder can contribute significantly to its highest residual resistance compared to that of other two samples [Fig. S6 in SM].

Further, the zero-field $R(T)$ measurements were performed with the excitation current (100 nA) much less than the critical current. Hence, the dissipation related to current-induced unbound vortex-antivorex movement can also be disregarded. In reduced dimension, phase fluctuation is one of the dominant mechanisms behind the resistive tailing & residual resistance[44] and above $T_{BKT}$, the phase fluctuation leads to vortex proliferation which eventually brings the system to the normal state[45]. In quasi 0D granular superconductors, depending on the dimension of the nano-grains, phase fluctuations can lead to total suppression of superconductivity and may lead to even an insulating state[46,47]. Besides, we have observed the PSLs for all the samples indicating significant contributions from phase fluctuations to the resistive states. Similarly, due to the nanoscale dimension of the grains, FSE can play the substantial role in the resistive tailing and the residual resistance[42,48] and also to the deviation of the BKT fit from the experimental data[17,38,42]. When the quantized energy level spacing due to FSE in quasi 0D nanoscale grains becomes comparable to the superconducting energy gap, residual resistance appear due to the decreased density of states at the Fermi level, Coulomb repulsion and suppression of Josephson coupling between the grains[35]. It is obvious that the impact of FSE is going to be the most for sample B3 with the smallest grain size among the samples. Indeed, it is observed in Fig. 7(f) (Fig. S6 in SM) that the residual resistance is the highest in B3. Therefore, we can conclude that the observed resistive tailing and the residual resistance in the nitride samples may be originated due to mainly the combined effect of disorder, quantum fluctuations and FSE.

## Conclusion:

We have demonstrated a simple technique to transform Nb into $Nb_2N$ by employing $Si_3N_4$ based substrate which serves as the source of nitrogen when it gets decomposed by high temperature annealing. The transformed nitride samples show unconventional superconductivity below 1K by exhibiting wide two step NM-SC transition featuring resistive tailing in the superconducting state. We emphasize here that the granularity mediated inhomogeneity, quantum fluctuation and the FSE are the main reasons behind the observed wide transition, the resistive tailing and the residual resistance in the *R(T)* characteristics. Interestingly, the current driven metal-superconductor transition as appeared in the *IVC*s also exhibits wide transition featuring stair-case type resistive steps that are the signatures of PSLs in 2D superconducting strips. The PSLs indicate a significant role of phase fluctuations to the transition too. Finally, our results demonstrate that $Nb_2N$ can be a promising candidate to study SFs and PS related phenomena and applications.

## Methods:

Nb thin films were grown on $Si_3N_4$/Si(100) [nitride] and $SiO_2$/Si(100) [oxide] substrates by using an ultra-high vacuum (UHV) DC magnetron sputtering system. For nitride substrate, low pressure chemical vapor deposition (LPCVD) grown 100 nm thick $Si_3N_4$ layer and for oxide substrate, thermally grown 300 nm thick $SiO_2$ layer act as the dielectric spacers to isolate the films from the substrates. Prior to deposition, substrates were gone through a rigorous cleaning process by ultrasonic cleaning in acetone and iso-propanol for 15 minutes in each. Afterwards, the substrates were cleaned in oxygen plasma for 15 minutes and finally, they were preheated at ~820°C in high vacuum (~1 × $10^{-7}$ Torr) inside the UHV chamber for 30 minutes to remove any chemical residues and surface contaminants. In the preheating stage, some parts of the substrate were covered with stainless steel shadow mask in order to have defined strips of Nb film. After the cleaning procedures we evacuated the chamber to less than 3.5 × $10^{-9}$ Torr and meanwhile the substrates were heated up to (820±10)°C which was maintained during the sputtering and post-sputtering annealing process. The sputtering was done with a Nb (99.99%) target in the presence of high purity Ar (99.9999% purity) gas at about $5 \times 10^{-3}$ mBar. The heating was continued after the deposition and post sputtering annealing was done at $(820\pm10)°C$ for 2 hours in high vacuum condition ($0.5\text{-}1.5 \times 10^{-7}$ Torr).

For low temperature transport study, electrical contact leads of Au(100 nm)/Ti(5 nm) were defined and aligned by a complimentary set of shadow mask on top of Nb strips. The length between the voltage probes for all the measured samples was ~1.1mm. The transport measurements were done in conventional 4-probe geometry with 100 nA excitation current by using ac lock-in technique in a dilution refrigerator (Triton from Oxford Instruments). For Structural characterization we have used grazing incidence X-ray diffraction (GIXRD) technique by Philips X'pert pro X-ray diffractometer using Cu-kα radiation operating at 40 kV and 20mA.

34   Zhang, G. *et al.* Global and Local Superconductivity in Boron-Doped Granular Diamond. *Adv. Mater.* **26**, 2034-2040 (2014).

35   Bose, S. & Ayyub, P. A review of finite size effects in quasi-zero dimensional superconductors. *Rep. Prog. Phys.* **77**, 116503 (2014).

36   Carbillet, C. *et al.* Confinement of superconducting fluctuations due to emergent electronic inhomogeneities. *Phys. Rev. B* **93**, 144509 (2016).

37   Yonezawa, S., Marrache-Kikuchi, C. A., Bechgaard, K. & Jérome, D. Crossover from impurity-controlled to granular superconductivity in (TMTSF)$_2$ ClO$_4$. *Phys. Rev. B* **97**, 014521 (2018).

38   Lin, Y.-H., Nelson, J. & Goldman, A. M. Suppression of the Berezinskii-Kosterlitz-Thouless Transition in 2D Superconductors by Macroscopic Quantum Tunneling. *Phys. Rev. Lett.* **109**, 017002 (2012).

39   Xu, C., Wang, L., Liu, Z., Chen, L., Guo, J., Kang, N., Ma, X.-L., Cheng, H.-M. & Ren, W. Large-area high-quality 2D ultrathin Mo2C superconducting crystals. *Nat. Mater.* **14**, 1135 (2015).

40   Benfatto, L., Castellani, C. & Giamarchi, T. Broadening of the Berezinskii-Kosterlitz-Thouless superconducting transition by inhomogeneity and finite-size effects. *Phys. Rev. B* **80**, 214506 (2009).

41   Caprara, S., Grilli, M., Benfatto, L. & Castellani, C. Effective medium theory for superconducting layers: A systematic analysis including space correlation effects. *Phys. Rev. B* **84**, 014514 (2011).

42   He, Q. L. *et al.* Two-dimensional superconductivity at the interface of a Bi2Te3/FeTe heterostructure. *Nat. Commun.* **5**, 4247 (2014).

43   Andersson, A. & Lidmar, J. Scaling, finite size effects, and crossovers of the resistivity and current-voltage characteristics in two-dimensional superconductors. *Phys. Rev. B* **87**, 224506 (2013).

**Acknowledgments:**

We gratefully acknowledge Dr. Sudhir Husale for critical reading of the manuscript and providing his invaluable comments and suggestions. The technical support for surface morphology imaging by AFM using the central facilities at CSIR-NPL is highly acknowledged. We are thankful to Mr. M. B. Chhetri for his technical assistance in the UHV sputtering lab. S.Y. acknowledges to UGC for JRF fellowship. B.G. and D.S. acknowledges to UGC-RGNF for SRF and JRF fellowships respectively. Authors acknowledge the financial support for establishing the dilution refrigerator facility at CSIR-NPL from the Department of Science and Technology (DST), Govt. of India, under the project, SR/S2/PU-0003/2010(G). This work was supported by CSIR network project 'AQuaRIUS' (Project No. PSC 0110) and is carried out under the mission mode project "Quantum Current Metrology".

## Author contributions:

B.G., S.Y., D.S. and S.S. fabricated the devices. B.G., S.Y., R.P.A. and S.S. carried out the low temperature measurements in dilution refrigerator. A.G. supported the transport measurements in dilution refrigerator. K.K.M. performed the XRD characterization. B.G., S.Y. and S.S. analyzed the data. S.S. planned, designed and wrote the manuscript and supervised the project. All the authors read and reviewed the manuscript.

## Additional information

**Supporting Material** contains the following,

1. AFM morphology for nitride samples B1, B2 and B3 and the Nb control samples $B1^{Ox}$, $B2^{Ox}$ and $B3^{Ox}$.
2. *IVC*s for oxide samples $B_1^{Ox}$, $B_3^{Ox}$ for both up and down current sweep directions.
3. *IVC*s for both up and down current sweep directions for nitride samples B1 and B2.
4. *R(T)* measurements in presence of perpendicular magnetic field for nitride samples B1, B2 and B3.
5. Calculation of Ginzburg-Landau (GL) coherence length ($\xi_{GL}$) for B1, B2 and B3.
6. Resistive tailing in the SC-state.
7. Summary Table: Characteristic parameters for the nitride samples.

**Competing interests:** The authors declare no competing financial and/or non-financial interests in relation to the work described.

**Figure Captions:**

**Fig. 1:** (a)-(d) Schematic presentation of the nitridation process for the transformation of Nb to $Nb_2N$ by using high temperature annealing with $Si_3N_4$/Si substrate. Arrows indicate the step wise process protocol. (e) Chemical representation of the reaction between Nb and $Si_3N_4$ under high temperature annealing leading to the formation of $Nb_2N$.

**Fig. 2:** XRD characterization. A set of XRD spectra for 4 nitride samples varying in thickness and one control sample of Nb film on $SiO_2$/Si substrate. With decreasing thickness for the nitride substrates, elemental Nb phase disappears and $Nb_2N$ phase starts to dominate. The samples G3 and G4 are the reference samples grown simultaneously with B1 and B2, respectively, where B1 and B2 have been used to study the transport characteristics presented in the subsequent section.

**Fig. 3:** Temperature dependent resistance [$R(T)$] measured at zero-field. (a) A set of for three nitride samples with varying thickness and three related to Nb control samples. For each thickness, one nitride sample and one control sample from the same batch have been selected. Inset: device geometry. (b) The same set of $R(T)$ is shown with the resistance values normalized by the normal state resistance ($R_N$) of individual sample. (c) Normalized resistance for the nitride samples are presented in reduced temperature ($T/T_{c0}$) scale. The vertical dotted lines represent the two transition regimes region-I and region-II. (d) Variation of different transition temperature ($T_c$) values, defined in (b), with thickness for the nitride

samples. The shaded regions indicate the dependence of transition width for the two transition regions on the sample thickness.

**Fig. 4:** Current-voltage characteristics (*IVC*s) of Nb/Si$_3$N$_4$ samples measured under zero-field condition. *IVC* isotherms for (a) B1, (b) B2 and (c) B3. (d) The *IVC*s for B3 are shifted in voltage for clarity. The steps in the *IVC*s are extended by the dotted lines to show their convergence at a current, known as excess current $I_s$ which is lower than the critical current. The existence of the excess current and the increasing slopes for the intermediate steps for increasing current direction (up sweep) are the signature of phase slip lines. Critical currents are defined as shown by the dotted arrows. The line arrows indicate the current sweeping direction.

**Fig. 5:** Temperature dependence of the characteristic currents for samples B1 (a), B2 (b) and B3(c). The temperature dependent critical currents and retrapping currents for the control samples on oxide substrates B1$^{ox}$ and B3$^{ox}$ are presented in the insets of (a) and (c), respectively. (d) Comparison of the critical currents between all the three nitride samples. Here, the critical temperature $T_C$ is defined as the temperature at which *IVC* becomes linear.

**Fig. 6:** *IVC*s for sample B1 under perpendicular magnetic field measured at 60 mK.

**Fig. 7:** Representation of NM-SC phase transition of Nb$_2$N thin films pictographically by using *R(T)* characteristics and the granular structure of the samples. (a) Normalized resistance in reduced temperature scale separately for all the three nitride samples. The colored shaded regions bounded by the dotted vertical

lines represent different stages of the phase transition and the stages are illustrated schematically through (b) to (e). Illustration of the granular films in normal state (b), establishment of local superconductivity in individual grains for the transition region-I (c), formation of superconducting puddle by combining phase coherent superconducting nano-grains in region-II (d), and finally the establishment of global coherence among the puddles to achieve the superconducting state (e). The solid cyan curves shown in (a) for the resistive transition in region-II represent the BKT transition using Halperin-Nelson equation which provides the BKT transition temperature $T_{BKT}$ and the fittings for individual samples are shown in (f) which selectively displays the fitting regions for all the three nitride samples. The details are explained in the text.

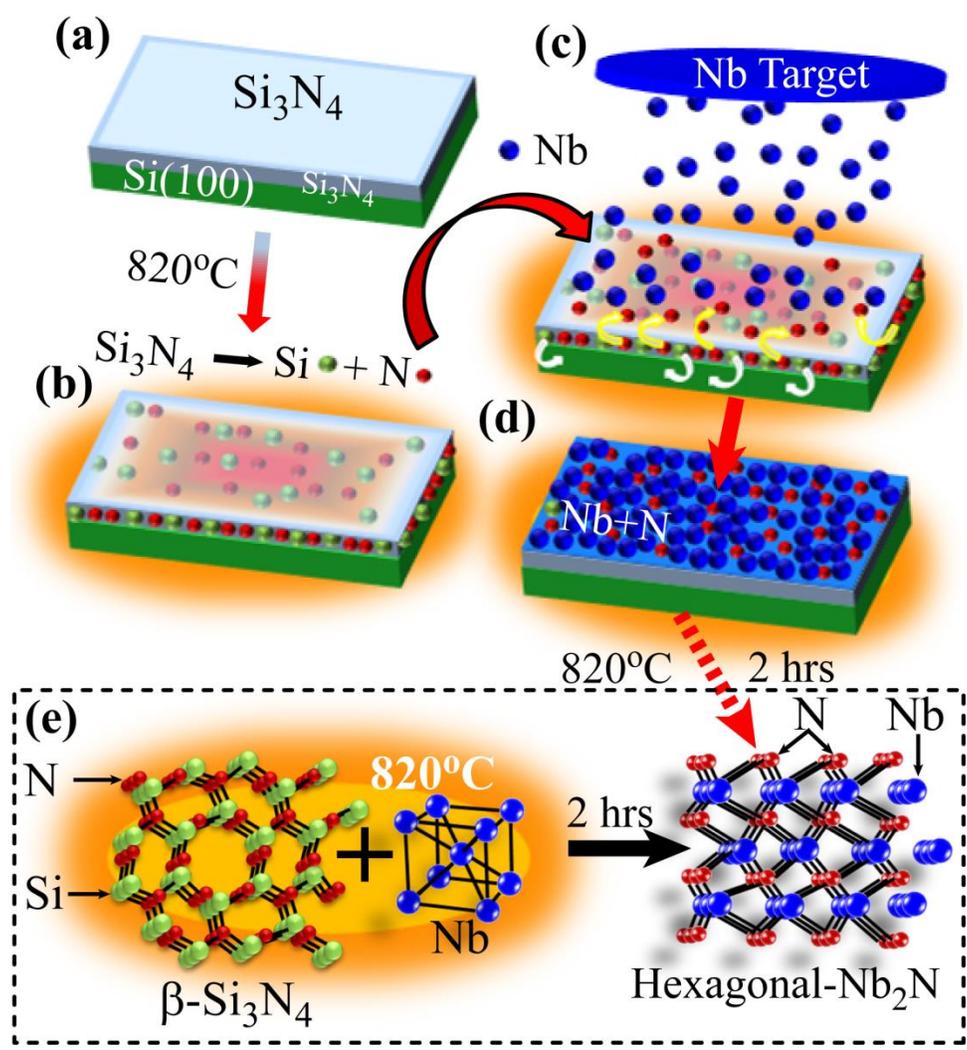

**Fig. 1**

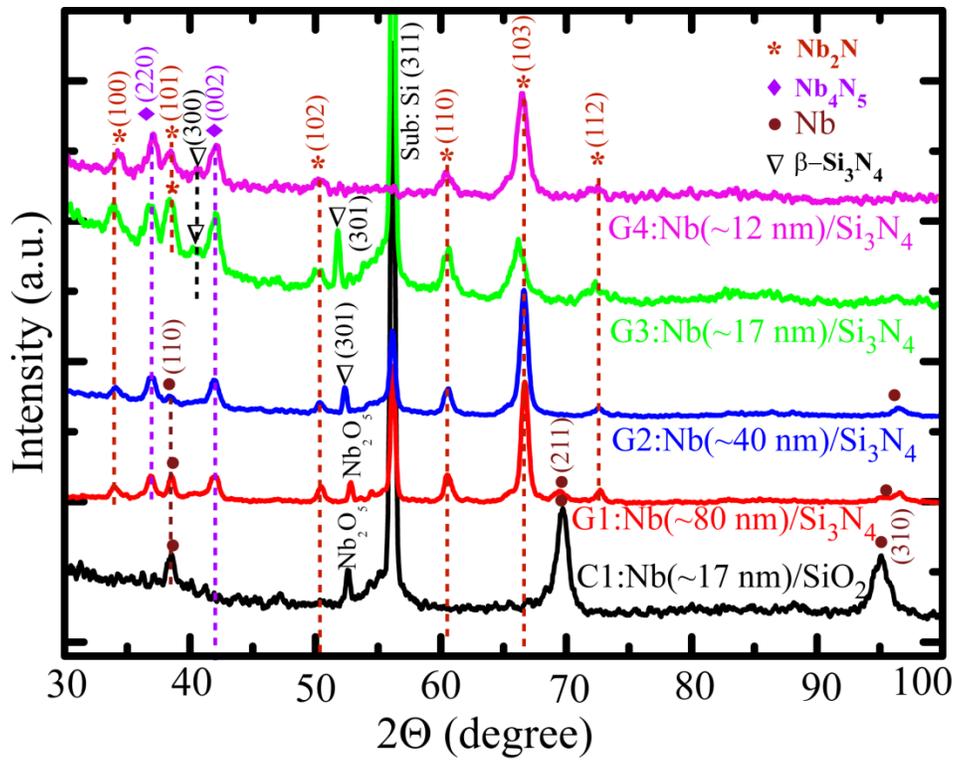

**Fig. 2**

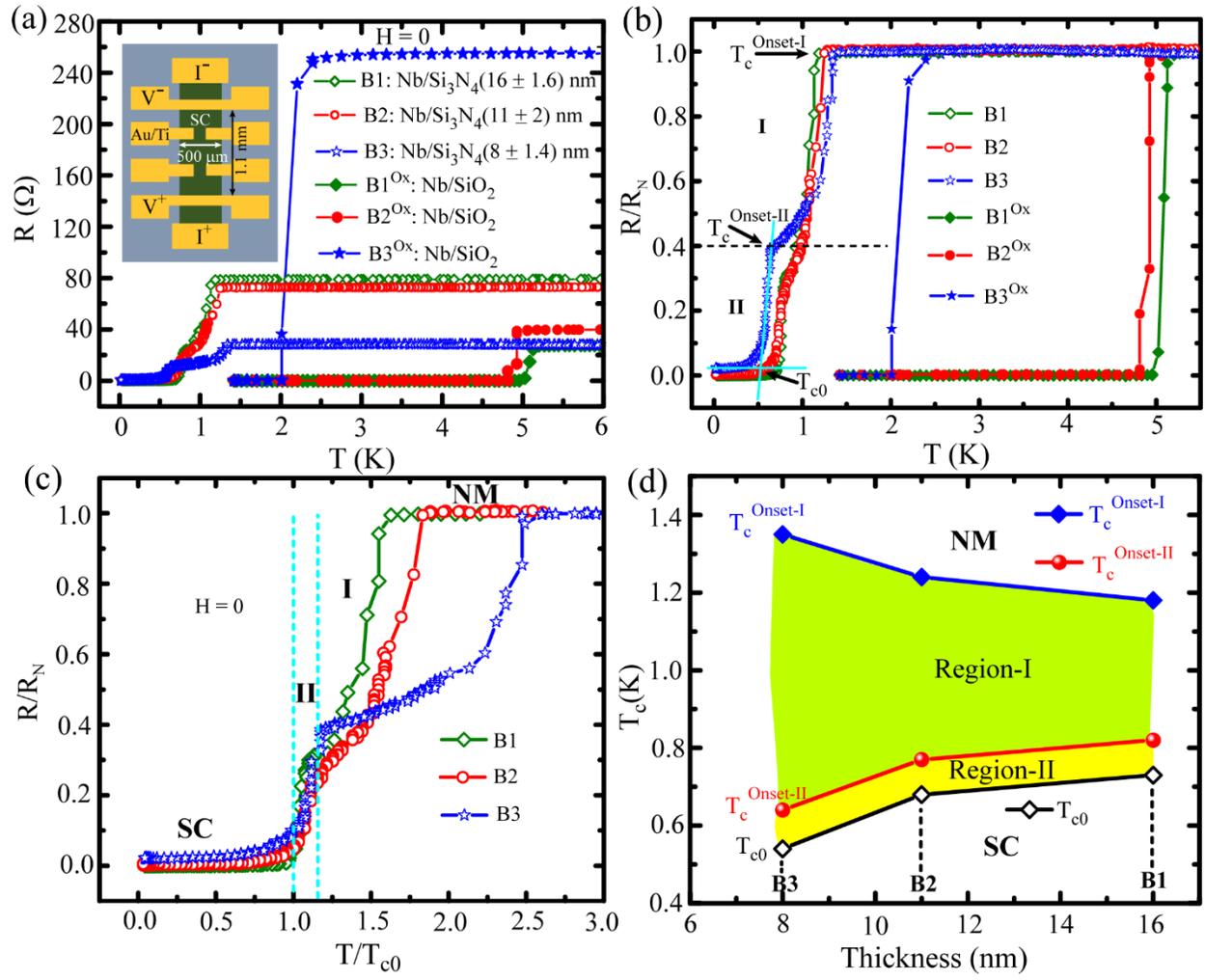

**Fig. 3**

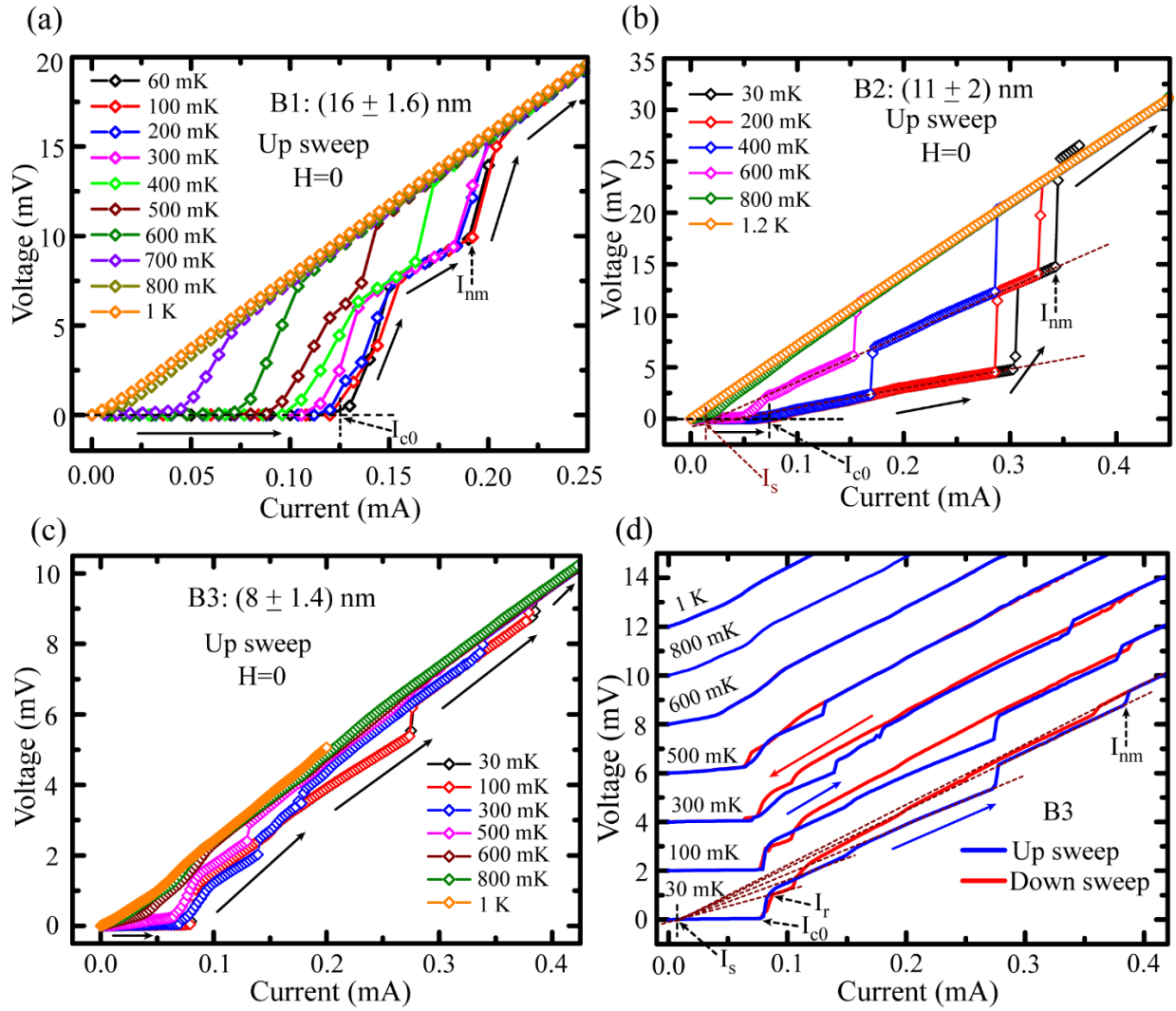

**Fig. 4**

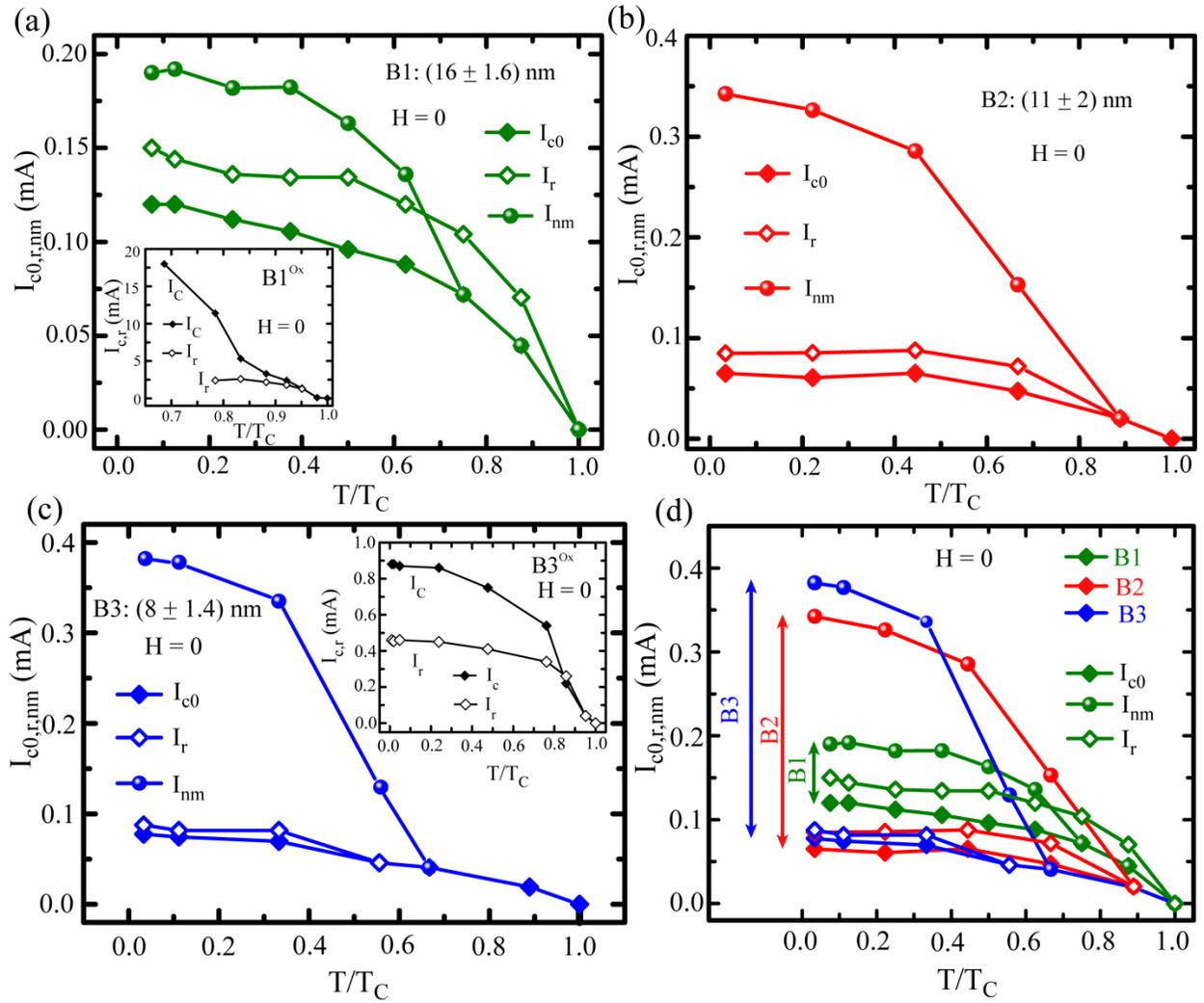

**Fig. 5**

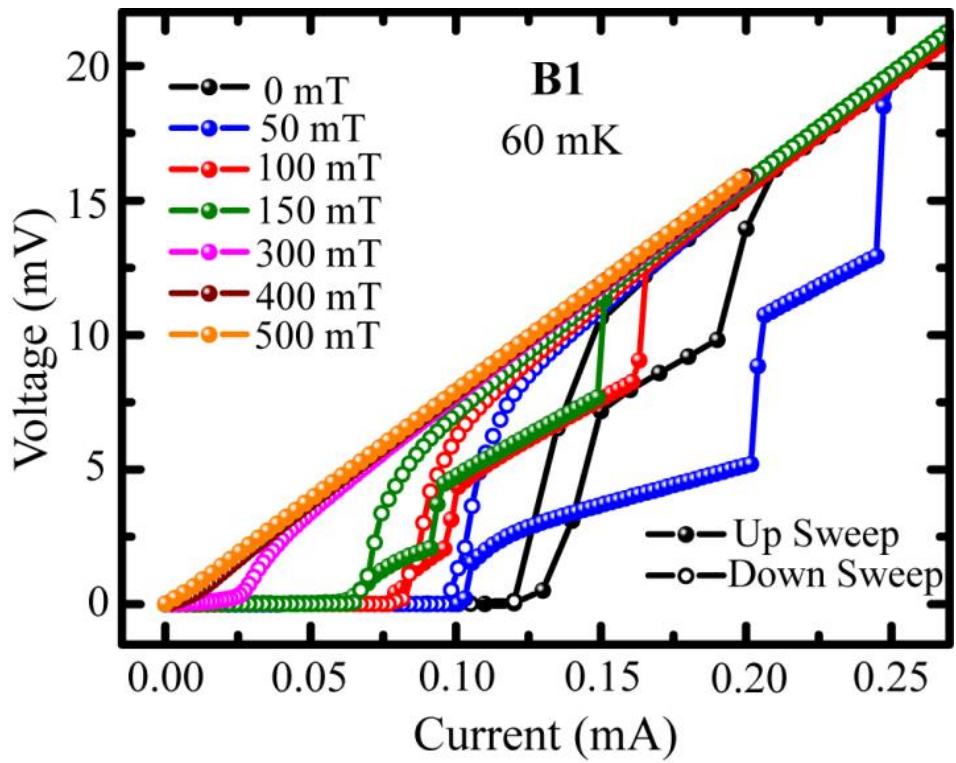

**Fig. 6**

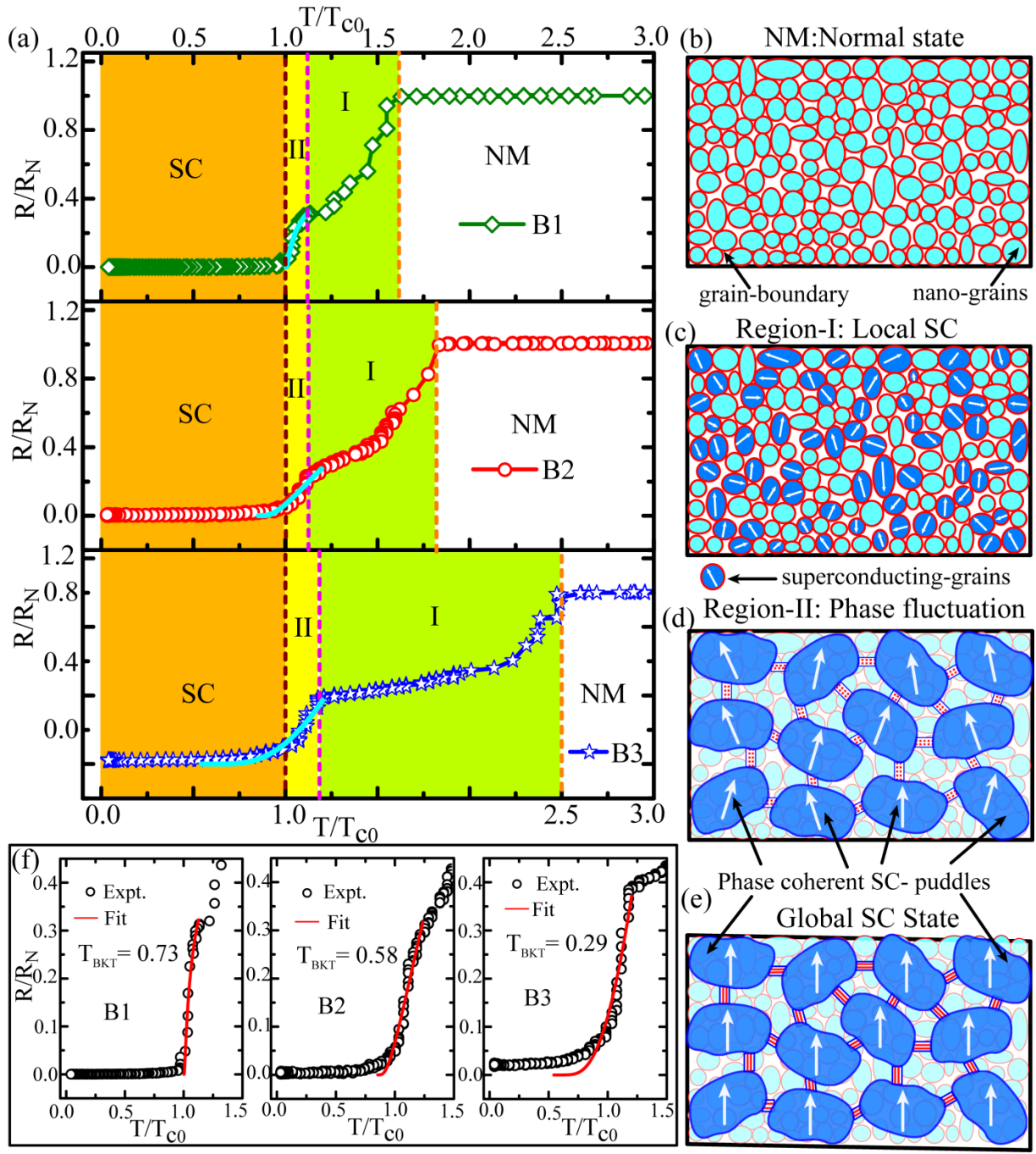

**Fig. 7**

## *Supporting Material*

# Substrate mediated nitridation of niobium into superconducting Nb$_2$N thin films for phase slip study

*Bikash Gajar[1,2], Sachin Yadav[1,2], Deepika Sawle[1,2], Kamlesh K. Maurya[1,3], Anurag Gupta[1,2], R.P. Aloysius[1,2], and Sangeeta Sahoo*[1,2]*

[1] *Academy of Scientific and Innovative Research (AcSIR), AcSIR Headquarters CSIR-HRDC Campus, Ghaziabad, Uttar Pradesh 201002, India.*

[2] *Electrical & Electronics Metrology Division, National Physical Laboratory, Council of Scientific and Industrial Research, Dr. K. S Krishnan Road, New Delhi-110012, India.*

[3] *Indian Reference Materials Division, National Physical Laboratory, Council of Scientific and Industrial Research, Dr. K. S Krishnan Road, New Delhi-110012, India.*

*\*The correspondence should be addressed to S.S. (Email: sahoos@nplindia.org)*

**Contents:**

1. **AFM morphology for nitride samples B1, B2 and B3 and the Nb control samples B1$^{Ox}$, B2$^{Ox}$ and B3$^{Ox}$.**

2. *IVCs* **for oxide samples B1$^{Ox}$, B3$^{Ox}$ for both up and down current sweep directions.**

3. *IVCs* **for both up and down current sweep directions for nitride samples B1 and B2.**

4. *R(T)* **measurements in presence of perpendicular magnetic field for nitride samples B1, B2 and B3.**

5. **Calculation of Ginzburg-Landau (GL) coherence length ($\xi_{GL}$) for B1, B2 and B3.**

6. **Resistive tailing in the SC-state**

7. **Summary Table: Characteristic parameters for the nitride samples.**

1. **AFM morphology for nitride samples B1, B2 and B3 and the Nb control samples B1$^{Ox}$, B2$^{Ox}$ and B3$^{Ox}$:**

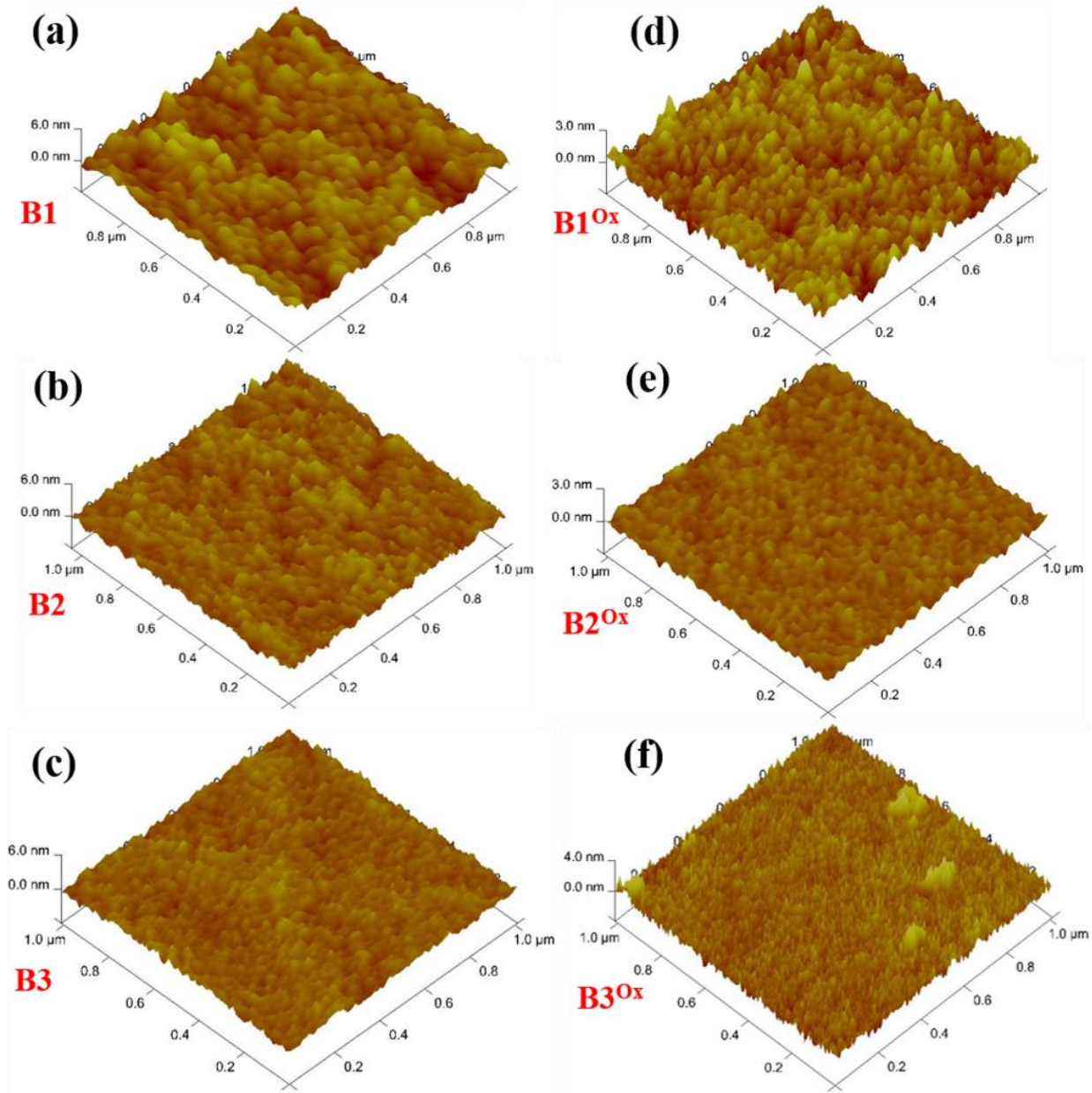

**Fig. S1: Morphological characterization. AFM topography images showing the granular nature of the films with variations in grain sizes for nitride samples B1, B2 and B3 in (a), (b) and (c), respectively and for the Nb control samples B1$^{Ox}$, B2$^{Ox}$ and B3$^{Ox}$ in (d), (e) and (f), respectively.**

AFM topography images, presented in Fig. S1, show the surface morphology of the nitride samples B1, B2 and B3 and the Nb control samples B1$^{Ox}$, B2$^{Ox}$ and B3$^{Ox}$. Both the nitride samples and the Nb control samples look granular in nature. The average grain sizes are estimated to be 40 nm, 30 nm and 23 nm for the nitride samples B1, B2 and B3, respectively. The overall variation in the grain size for the region bounded by the voltage probes are about ± 10 nm for all the three samples. Therefore, grain size reduces for reduced thickness. Further, the surface looks less uniform for the thinner samples than that appeared for the thicker one. On the other hand, the estimated average grain sizes are 37 nm, 27 nm and 12 nm for the control samples B1$^{Ox}$, B2$^{Ox}$ and B3$^{Ox}$, respectively with an overall variations of ± 7 nm for all of them. Grain sizes follow the same detrimental trend for the control samples also with reduction in thickness. For control samples, the reduction in $T_c$ can be understood by the grain size reduction and for the thinnest sample B3$^{Ox}$ the grain size is much lower if we compare the same for the thinnest nitride sample B1. However, we do not observe staircase like phase slip lines (*PSLs*) in the current voltage characteristics (*IVCs*) for B3$^{Ox}$ and the *R(T)* was observed to be sharp with no such resistive tailing which was present for all the nitride samples.

## 2. *IVCs* for oxide samples $B1^{Ox}$, $B3^{Ox}$ for both up and down current sweep directions:

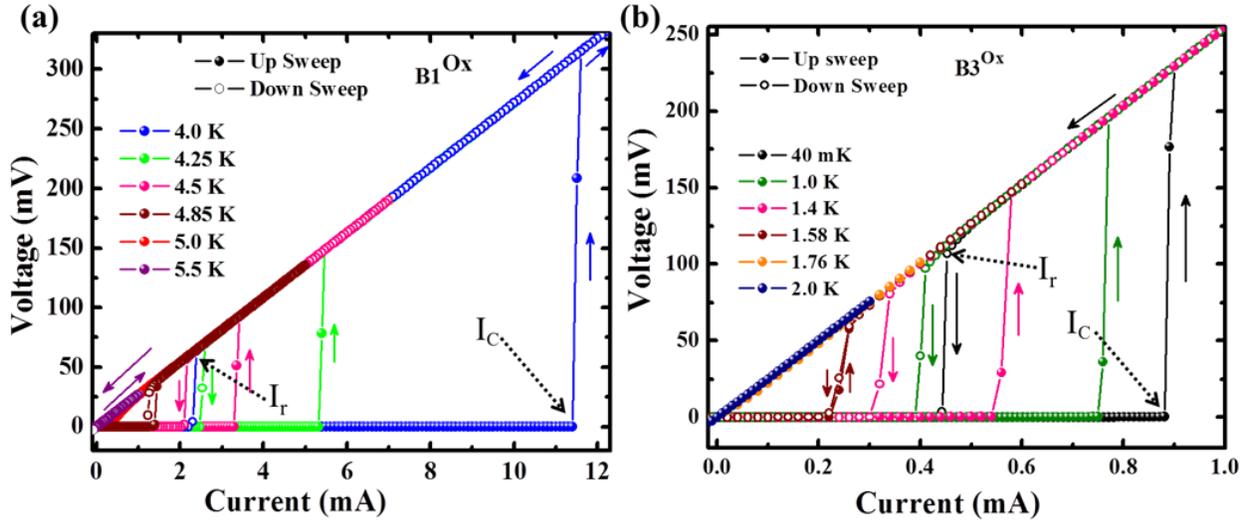

**Fig. S2: Current voltage characteristics (IVCs) for oxide samples. IVC isotherms for both up and down current sweep directions for $B1^{Ox}$ (a) and $B3^{Ox}$ (b). The critical current ($I_C$) and the retrapping current ($I_r$) are marked by the dotted arrows while the solid arrows show the sweep direction for the bias current.**

In Fig. S2, we have shown the *IVCs* for the oxide samples for both up and down sweep directions for the bias current. Here we selected the thickest ($B1^{Ox}$) and the thinnest ($B1^{Ox}$) samples among the three samples in order to observe the effects of thinning on the *IVCs* for the oxide samples. As we have already seen in the main text that for the nitride samples thickness plays a vital role in monitoring the transition region for superconductor-metal transition in their *R(T)* and also in *IVCs*. However, for the oxide samples we observe a single step sharp transition from superconductor to normal metallic transition at the critical current $I_C$. The same is observed in the reverse direction at the retrapping current $I_r$ which is much less than $I_C$. Hence, the *IVCs* are hysteretic with respect to the current sweeping direction. Except for the values of these two characteristic currents $I_C$ and $I_r$ both the samples show similar type of hysteretic *IVCs* which show a direct transition from SC to NM and *vice-versa*. It should be noted here that contrary to the nitride sample, we do not observe any intermediate resistive states for the oxide samples.

## 3. *IVCs* for both up and down current sweep directions for nitride samples B1 and B2:

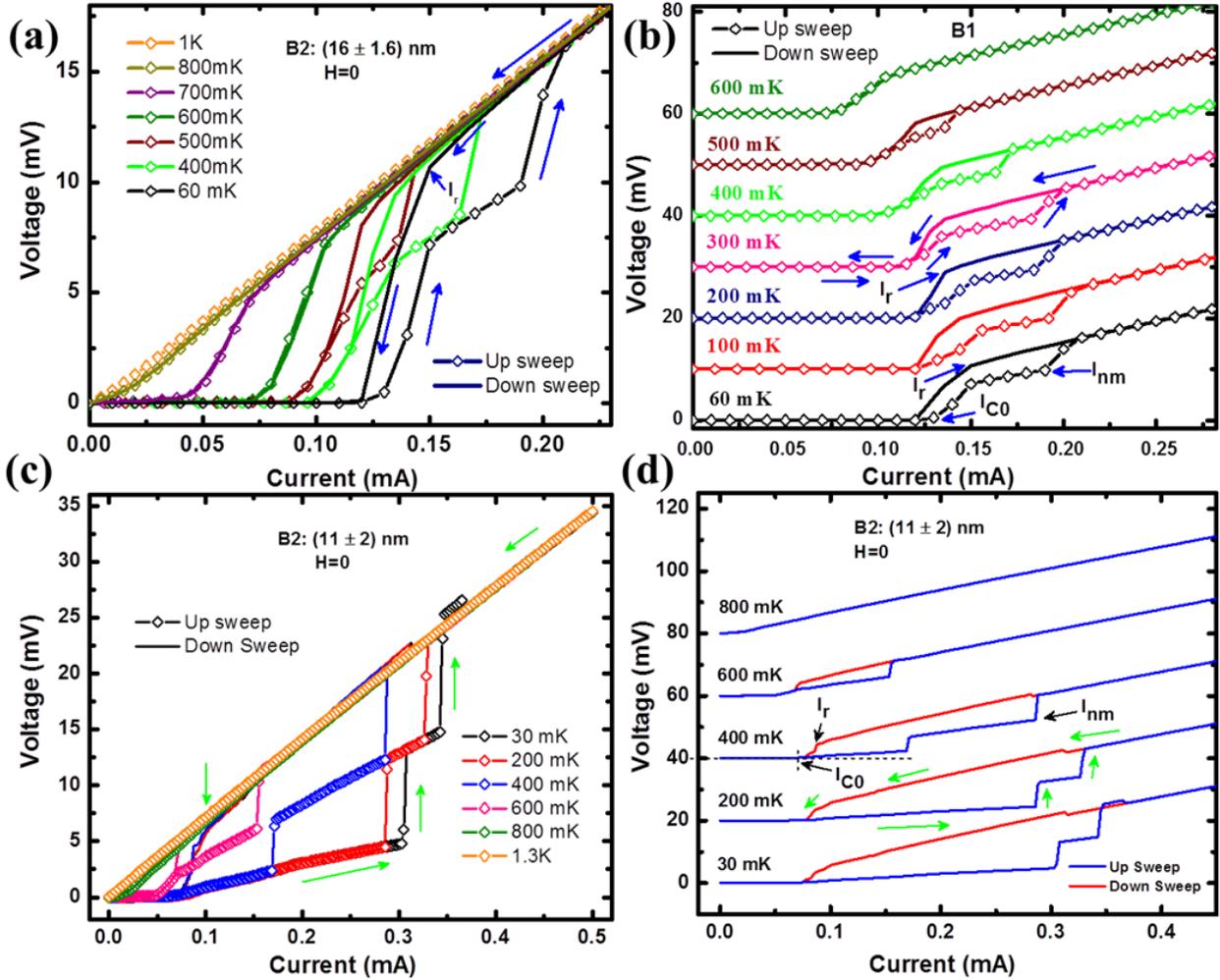

**Fig. S3:** Zero-field IVCs for nitride samples B1 and B2 for both increasing & decreasing current sweeping directions. Isothermal IVCs for B1 (a) and for B2 (c). For clarity, the IVC isotherms are shifted in the voltage axis by 10 mV from the consecutive IVC isotherm for B1 (b) and by 20 mV for B2 (d), respectively. The superconducting state in each IVC corresponds to zero-voltage in the voltage axes presented in (b) & (d). The characteristic currents, namely, the retrapping current $I_r$, the critical currents $I_{c0}$ & $I_{nm}$ are defined by the arrows in (b) and (d).

## 4. *R(T)* measurements in presence of perpendicular magnetic field for nitride samples B1, B2 and B3:

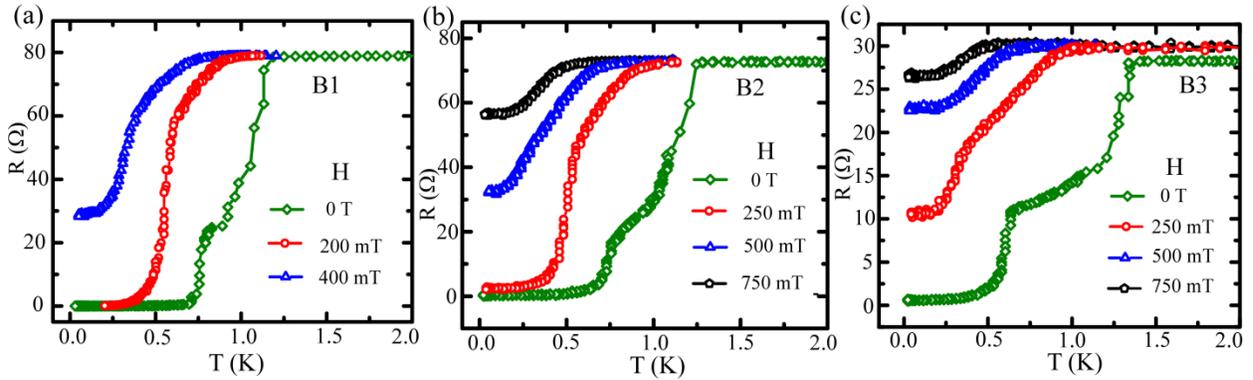

**Fig. S4: Effects of perpendicular magnetic field on R(T) characteristics for nitride samples. Field-dependent R(T) for B1 (a), B2 (b), and B3 (c), respectively.**

In the main text, we have discussed about zero-field *R(T)* measurements in detail for the nitride samples. In Fig. S4, we present field-dependent *R(T)* including the zero-field *R(T)* separately for all the three samples. Here the variations in the zero-field *R(T)* characteristics among the samples are clearly visible. The curvature in the region-I is getting much more prominent with reduction in thickness from samples from B1-to-B2 and from B2-to-B3. With application of an external magnetic field of about 200-250 mT, the curvature in region one gets almost reversed and the transition shifts towards lower temperature. The kink, separating the two regions as explained in detail in the main text, almost disappears with the application of field and the *R(T)* characteristics look smooth compared to that measured at zero—field. The curvature in region-II remains almost unchanged for all the samples when measured under the external field.

## 5. Calculation of Ginzburg-Landau (GL) coherence length ($\xi_{GL}$) for B1, B2 and B3:

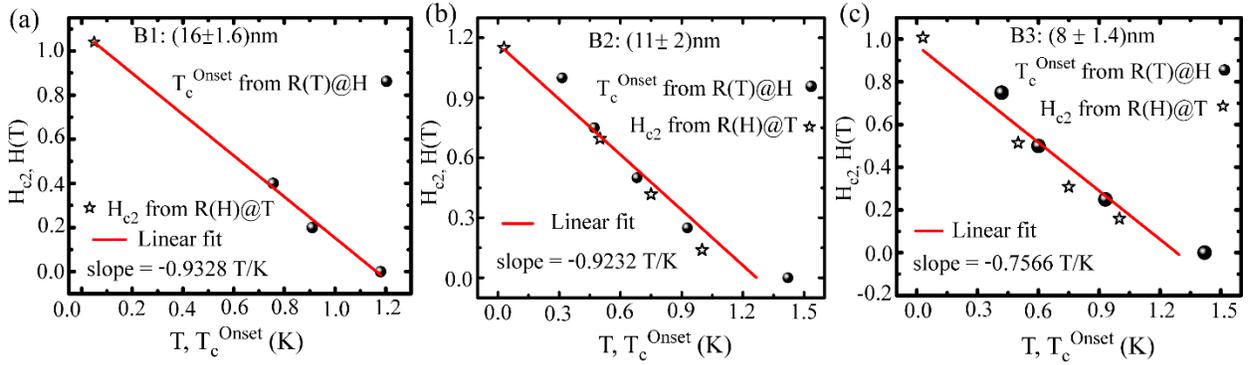

**Fig. S5: H-T phase diagram for nitride samples. Experimental points are obtained from magnetoresistance [R(H)] isotherms and from field dependent R(T) measurements under fixed external magnetic field applied perpendicular to the sample. The linear fit confirms that the samples are in dirty limit. The slope provides the GL coherence length $\xi_{GL}$, the details are explained**

We have calculated the Ginzburg-Landau coherence length, $\xi_{GL}(0)$, from the upper critical field $H_{c2}(Tesla)$ using the following formula[1,2], $\xi_{GL}(0) = \left[\dfrac{\phi_0}{2\pi T_c \left|\dfrac{dH_{c2}}{dT}\right|_{T_c}}\right]^{1/2}$, where $\phi_0$ is the flux quantum. In order to estimate $\xi_{GL}(0)$, we have plotted the temperature dependent $H_{c2}$ for all the three nitride samples in Fig. S5. We have obtained the values for $H_{c2}$ from isothermal magnetoresistance $R(H)$ measurements and the related points are marked as the star-shaped scattering points. Further, we have calculated $T_c^{Onset}$ from the field dependent $R(T)$ measurements carried out under a fixed external magnetic field and the points are represented by the solid spherical scattering points in Fig. S5. The experimental data from both *R(H)* and *R(T)* measurements follow a linear variation between the critical field and the critical temperatures. The slope of the fit is used to calculate the GL coherence length, $\xi_{GL}(0)$. Here, we used $T_c^{Onset-I}$ as the $T_c$ and accordingly, we have evaluated $\xi_{GL}(0)$ as 17.3 nm, 16.96 nm and 17.96 nm for B1, B2 and B3, respectively. As our thickness values for the nitride samples are of the order or less than the related coherence lengths, therefore, the samples can be considered in 2D limit.

## 6. Resistive tailing in the SC-state

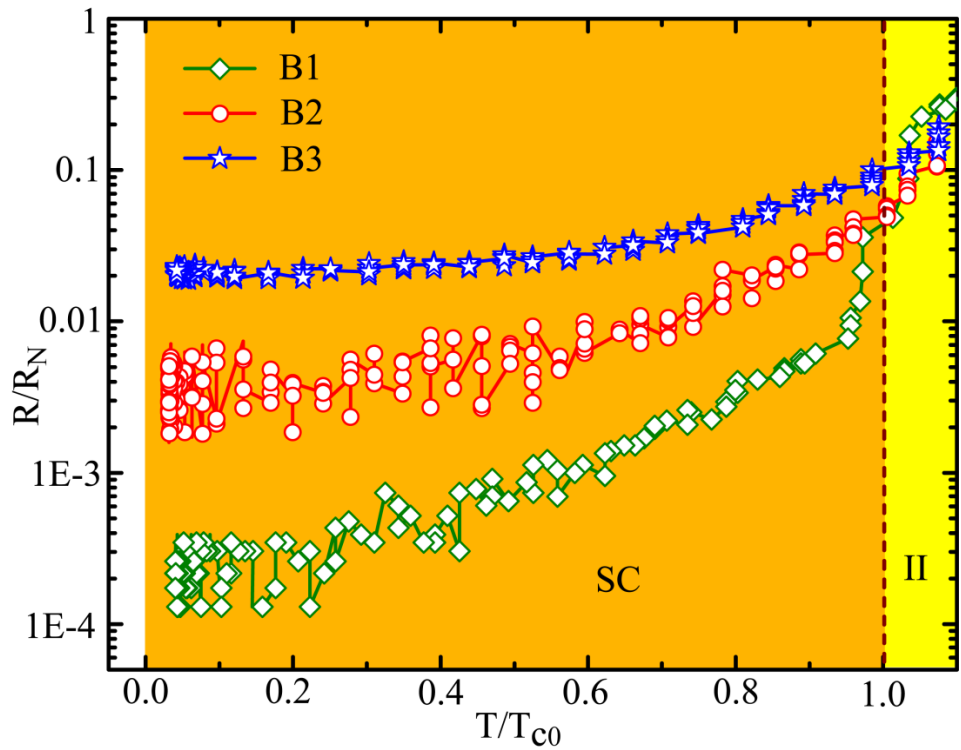

**Fig. S6: A collective representation of the SC-states in a semi-logarithmic plot to have the insight into the SC-state.**

## 7. Summary Table: Characteristic parameters for the nitride samples

**Table 1: Characteristic parameters for the nitride samples**

| Sample | Thickness (nm) | Grain size (nm) ±10 nm | $R_N$ ($\Omega$) | $T_c^{Onset-I}$ (K) | $T_c^{Onset-II}$ (K) | $T_{c0}$ (K) | $T_C^{IV}$ (K) | $\Delta T_I$ (K) | $\Delta T_{II}$ (K) | $\Delta T_{II}/T_{C0}$ | R@ $T_c^{Onset-II}$ ($\Omega$) |
|---|---|---|---|---|---|---|---|---|---|---|---|
| **B1** | 16 | 40 | 79 | 1.18 | 0.82 | 0.73 | 0.8 | 0.36 | 0.09 | 0.123 | **0.4 $R_N$** |
| **B2** | 11 | 30 | 72.3 | 1.24 | 0.77 | 0.68 | 0.9 | 0.47 | 0.09 | 0.132 | **0.25$R_N$** |
| **B3** | 8 | 23 | 28.3 | 1.35 | 0.64 | 0.54 | 0.9 | 0.71 | 0.1 | 0.185 | **0.3 $R_N$** |

We have summarized the characteristic parameters that are obtained from the *R(T)* and *IVC* measurements for the nitride samples. The thickness and the grain sizes are measured by atomic force microscopy (AFM). Here, $T_C^{IV}$ relates the transition temperature obtained from *IVC*s and it is mentioned in the main text as $T_C$. $\Delta T_I$ and $\Delta T_{II}$ correspond to the transition width in region-I and region-II respectively. It is clear that transition width is much wider in region-I than that in region-II. The normal state resistance $R_N$ and the onset critical temperature $T_c^{Onset-I}$ vary in an unconventional way with the thickness. Further, the $T_C^{IV}$ appears to be close to $T_c^{Onset-II}$ and from the *IVC*s, we have observed that below $T_C^{IV}$ PSLs appear for all the samples. This indicates that the phase fluctuations in region-II and SC state contribute significantly to the resistive transition.